\documentclass[showpacs,twocolumn,prl]{revtex4}

\usepackage{amssymb}
\usepackage{amsmath}
\usepackage{epsfig}

\newcommand{\be}{\begin{equation}}
\newcommand{\ee}{\end{equation}}
\newcommand{\beqn}{\begin{eqnarray}}
\newcommand{\eeqn}{\end{eqnarray}}

\newcommand{\Hrep}{{\cal H}_{\mathrm{rep}}}

\newcommand{\ab}{_{\alpha\beta}}

\newcounter{RB}

\begin{document}

\title{Free energy fluctuations in Ising spin glasses}
\author{T. Aspelmeier}
\affiliation{Department of Physics and Astronomy, University of
Manchester, Manchester M13 9PL, UK}
\author{M. A. Moore}
\affiliation{Department of Physics and Astronomy, University of
Manchester, Manchester M13 9PL, UK}
\date{\today}

\begin{abstract}
The sample-to-sample fluctuations of the free energy in finite-dimensional
Ising spin glasses are calculated, using the replica method, from higher order
terms in the replica number $n$. It is shown that the Parisi symmetry breaking
scheme does not give the correct answers for these higher order terms. A 
modified symmetry breaking scheme with the same stability is shown to resolve 
the problem.
\end{abstract}


\maketitle

The Parisi replica symmetry breaking scheme \cite{Parisi79,Parisi80a,
Parisi80b,MezardEtAl87} was a milestone in the study of spin glasses. It was
proposed as long ago as 1979 and has been extensively used ever since.  It is
the purpose of this Letter to point out that an important modification of it
is needed in order to calculate the sample-to-sample fluctuations of spin
glasses.

The replica method is usually used in spin glasses to calculate the average
free energy $F$ as a function of inverse temperature $\beta$ from the
partition function $Z$ via
\begin{align}
    \label{rep1}
\beta F &= -\overline{\ln Z} = -\lim_{n\to 0}\frac 1n \ln\overline{Z^n}.
\end{align}
The overbar means averaging over bond configurations. Apart from the free
energy, the replica method also gives in principle access to other physical
quantities as well. Expanding the logarithm on the right hand side in
Eq.~\eqref{rep1} one gets
\begin{align}
    \label{rep2}
    \ln\overline{Z^n} &=
    -n\beta F + \frac{n^2}{2}\beta^2 \Delta F^2 + \cdots ,
\end{align}
where $\Delta F^2=(\overline{\ln^2 Z}-\overline{\ln Z}^2)/\beta^2$ is the
mean-square sample-to-sample fluctuation of the free energy, and the
coefficients of the higher order terms are higher order cumulants. In order to
obtain the coefficients in this expansion from the replica method, it is
necessary to keep the replica number $n$ small but finite throughout the
calculation (as opposed to the case of the free energy itself where it is
possible to set $n=0$ early on), which makes it rather cumbersome. Moreover,
using Parisi's replica symmetry breaking scheme
\cite{Parisi79,Parisi80a,Parisi80b,MezardEtAl87}, it will be shown below that
the coefficients so obtained \textit{differ} when assuming $n$ positive or
negative. Note that the replica method intrinsically requires $n$ to be
non-integer, and that it is no more unnatural to take $n$ negative than it is
to take it non-integer. Thus conflicting answers are obtained for
the same physical quantity which indicates that there is a problem with  
Parisi's replica symmetry breaking scheme.
We will argue that the correct answer is in fact given by the $n<0$ solution.
Furthermore we show how the symmetry breaking scheme can be modified
to give the correct results when $n>0$. However, this new replica symmetry
breaking
scheme gives the same results as Parisi's scheme for quantities like the
free energy or the distribution of spin overlaps $P(q)$, and has the same
stability.

We start from the usual replica field theory for $d$-dimensional spin glasses
as derived, e.g., in \cite{BrayMoore79} or
\cite{DeDominicisEtAl98}, where we have a free energy functional
\begin{multline}
    \label{rep3}
    \Hrep\{q\ab\} = \int d^d x \, \left[
-{\tau \over 2} \sum_{\alpha,\beta} q\ab^2 +
{1 \over 4} \sum_{\alpha,\beta} (\vec{\nabla} q\ab)^2  \right. \\
\left.
-{w \over 6} \sum_{\alpha,\beta,\gamma}  q\ab q_{\beta\gamma}q_{\gamma\alpha} 
-{y \over 12} \sum_{\alpha,\beta} q\ab^4 
\right]
\end{multline}
and
\begin{align}
    \label{rep4}
    \overline{Z^n} &= \int\left(\prod_{\alpha<\beta} \mathcal{D}q\ab(x)\right)
      \exp(-\Hrep\{q\ab\}).
\end{align}
Here, $q_{\alpha\alpha}=0$, we have omitted some unimportant terms of order
$q^4$, and set $\tau = 1 - T/T_c$. The fourth order term included is the one
responsible for replica symmetry breaking. We assume that the dimension $d$ is
above the special dimension 8 to keep the loop expansion straightforward
\cite{DeDominicisEtAl98}.

Since we are keeping $n$ finite throughout, it will be necessary to review
some of the results that have been derived in \cite{Kondor83,%
TemesvariEtAl94,DeDominicisEtAl94,DeDominicisEtAl97,DeDominicisEtAl98} for
integer $n$, allowing us to write, e.g., $q\ab$ with integer indices $\alpha$
and $\beta$, and of results that have been derived for arbitrary (noninteger)
$n$ and with an infinite number of symmetry breaking steps, such that we must
re-write $q\ab$ in terms of $q(x)$ where $x$ is a continuous variable
between $n$ and $1$. We will
have to switch between these notations frequently but note that they are
equivalent. In order to distinguish better between the cases $n>0$ and $n<0$,
we will from now on use the notation $n^+$ for positive $n$ and $n^-$ for
negative $n$. The letter $n$ itself will be used when no distinction is
needed.

To Gaussian order we get from Eq.~\eqref{rep4}
\begin{align}
    \label{rep5}
    \ln\overline{Z^n} &= -\Hrep \{ q^{\text{SP}}\ab \} -
    \frac V2 \int\frac{d^dk}{(2\pi)^d}\sum_{\mu}d_{\mu}\ln(k^2+\lambda_{\mu}),
\end{align}
where $q^{\text{SP}}\ab$ is the Parisi saddle point solution, $k$ is a
$d$-dimensional wave vector and $\lambda_{\mu},d_{\mu}$ are the eigenvalues of
the Hessian, evaluated at the saddle point solution, and their degeneracies.

\begin{figure}
\epsfig{file=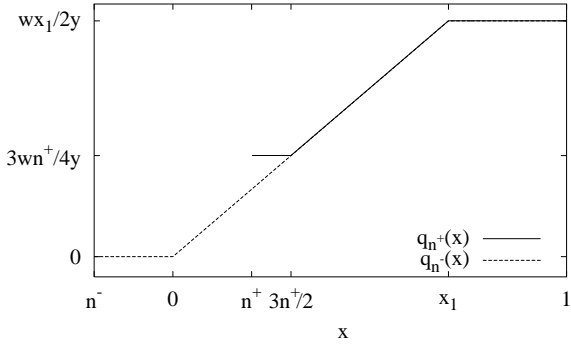,width=\columnwidth}
\caption{The two types of $q$-function for positive ($n^+$) and negative
($n^-$) replica number. For $x\ge 3n^+/2$, the two curves coincide.}
\label{qfunction}
\end{figure}

The first (mean field) term in Eq.~\eqref{rep5} has been worked out by Kondor
\cite{Kondor83} for finite $n^+$. We repeat his calculation here
for $n^-$ since it gives a qualitatively different result which we will need
later on. The free energy at mean-field level for the Parisi function 
$q_{n}(x)$ at finite
$n$ is (cf. Eq.~(9) of \cite{Kondor83})
\begin{multline}
    -\frac{\Hrep}{n} = \frac\tau2 \int_{n}^1 dx\,q_{n}^2(x) +
    \frac{y}{12}\int_{n}^1 dx\,q_{n}^4(x) - \\
    \frac w6\int_{n}^1 dx\,
    \left(xq_{n}^3(x) + 3q_{n}^2(x)\int_{x}^1 dt\,q_{n}(t)\right),
\end{multline}
when extremized with respect to $q_{n}(x)$. The solutions for 
$n^+$ and $n^-$ are
\begin{align}
    \label{qpos}
    q_{n^+}(x) &= \left\{
    \begin{array}{l@{\hspace{2em}}l}
      3wn^+/4y & n^+\le x \le 3n^+/2 \\
      wx/2y & 3n^+/2 < x \le x_{1} \\
      wx_{1}/2y & x_{1} < x \le 1
    \end{array}\right. \\
    \label{qneg}
    q_{n^-}(x) &= \left\{
    \begin{array}{l@{\hspace{2em}}l}
      0 & n^-\le x \le 0 \\
      wx/2y & 0 < x \le x_{1} \\
      wx_{1}/2y & x_{1} < x \le 1
    \end{array}\right. ,
\end{align}
where $x_{1}=1-\sqrt{1-4y\tau/w^2}$ is the usual breakpoint of the Parisi
$q$-function. Fig.~\ref{qfunction} shows the two solutions for
illustration.

We note that the kind of problem we are going to encounter already shows up at
this level: $\Hrep[q_{n^-}(x)]$ has \textit{no} terms of higher order than
$n^-$, while $\Hrep[q_{n^+}(x)]$ does have additional terms of order ${n^+}^6$
and higher \cite{Kondor83}. We also note that at mean-field level, the
variance $\Delta F^2$ is 0 since there is no $n^2$-term.

In order to calculate the fluctuation corrections, it is useful to define
\begin{align}
    \label{Idef}
    I &= \sum_{\mu}d_{\mu}\ln(k^2+\lambda_{\mu}).
\end{align}
By differentiating $I$ with respect to $k^2$ one obtains
\begin{align}
    \frac{\partial I}{\partial (k^2)} &=
    \sum_{\mu}\frac{d_{\mu}}{k^2+\lambda_{\mu}} 
    = \sum_{\alpha<\beta}G_{\alpha\beta,\alpha\beta},
\end{align}
where $G_{\alpha\beta,\gamma\delta}$ are the propagators, which are
essentially the inverse of the Hessian. The propagators have been calculated
exactly by De Dominicis et al. \cite{DeDominicisEtAl98} in the ``continuum
limit'' of infinitely many replica symmetry breaking steps. While their
results are too long to be quoted here, we mention that the diagonal
propagators $G_{\alpha\beta,\alpha\beta}$ are in this limit denoted by
$G^{xx}_{11}$, are labelled by a continuous variable $x\in[n,1]$, and we have
\begin{align}
    \label{propsum}
    \sum_{\alpha<\beta}G_{\alpha\beta,\alpha\beta} &= -\frac n2
    \int_{n}^1 dx\,G^{xx}_{11}.
\end{align}
For ease of notation, we will from now on drop the subscript $11$ from the
propagators (which is superfluous for our purposes) and replace it by $n$ to
indicate that the propagators here still depend on it. While the propagators
in \cite{DeDominicisEtAl98} were derived for $n=0$, it is possible but tedious
to extend the calculation to finite $n$. Fortunately, there is a simple
argument to obtain the propagators's exact form for $n^-$. Since they are
labelled by the variable $x$, which is nothing but the inverse of the Parisi
function, i.e.\ $x(q)$, they are effectively labelled by $q$. From this we can
infer that $G^{xx}_{n^-}=\text{const.}$ for those $x$ where $q_{n^-}(x)$ is a
constant, therefore (cf. Eq.~\eqref{qneg}) $G^{xx}_{n^-}=G^{00}_{n^-}$ for
$x\le 0$, and for $x\ge 0$, $G^{xx}_{n^-}=G^{xx}_{0}$ because
$q_{n^-}(x)=q_{0}(x)$. We can thus express the propagators for $n^-$ entirely
in terms of the $n=0$ propagators.  For $n^+$ this simple argument does not
work because $q_{n^+}(x)$ from Eq.~\eqref{qpos} is (almost) identical to the
$q$-function in a field, and it has been shown in \cite{DeDominicisEtAl94}
that this leads to an additional $n^+$-dependent term in the propagators. We
will see below that luckily we do not need to work this out in detail.

We are now in a position to calculate $\Delta F^2$ for $n^-$. First, we
calculate
\begin{align}
    \label{propint}
    \frac{\partial I}{\partial (k^2)} &=
    -\frac{n^-}{2} \int_{n^-}^1 dx\,G^{xx}_{n^-} =
    -\frac{n^-}{2} \int_{0}^1 dx\,G^{xx}_{0} + \frac{{n^-}^2}{2} G^{00}_{0}.
\end{align}
Thus there are no terms of higher order than ${n^-}^2$. This already implies,
by comparison with Eq.~\eqref{rep2}, that the free energy fluctuations have a
Gaussian distribution since all higher order cumulants are zero. In order to
calculate the variance of this distribution, $\Delta F^2$, we only need to
know the coefficient of the ${n^-}^2$-term, which is simply $G^{00}_{0}$, and
integrate it with respect to $k^2$. The propagator $G^{00}_{0}$ is given by
\cite{DeDominicisEtAl98}
\begin{multline}
    G^{00}_{0} = 
    \int_{0}^1\frac{ds}{s}\int_{0}^1\frac{dt}{t}
    \frac{\partial^2}{\partial s\partial t} f(s,t) \\
    - \int_{0}^1\left(\frac{ds}{s}\frac{\partial}{\partial s} f(s,x_1) +
    \frac{dt}{t}\frac{\partial}{\partial t} f(x_1,t)\right) + f(x_1,x_1),
\end{multline}
where
\begin{align}
    \label{fdef}
    f(s,t) &= \frac{1}{k^2+yq_{0}^2(s)+yq_{0}^2(t)}
\end{align}
is the inverse of the $x=0$ eigenvalues of the Hessian from the replicon
sector \cite{DeDominicisEtAl98}.  The function $f$ can easily be integrated
with respect to $k^2$\footnote{Expanding $I$ from Eq.~\eqref{Idef} for large
$k^2$ shows that it behaves like $\ln k^2 + \mathcal{O}(1/k^2)$, i.e.\ there
is no constant term of $\mathcal{O}(1)$. This property determines the constant
of integration and amounts to using the indefinite integral $\int d(k^2)\,f =
\ln(k^2+yq_{0}^2(s)+yq_{0}^2(t))$.}, and the integrals in $G^{00}_{0}$ can
then be worked out, resulting in
\begin{align}
    J &:= \int d(k^2)G^{00}_{0} \\
\begin{split}
    &= \ln(k^2+\frac{x_{1}^2w^2}{2y}) \\
&\quad - \frac{4w(4yk^2 + wx_{1})}{4yk^2\sqrt{4yk^2+w^2x_{1}^2}}
     \tan^{-1}\frac{wx_{1}}{\sqrt{4yk^2+w^2x_{1}^2}},
\end{split}
\end{align}
such that
\begin{align}
    \label{deltaf}
    \beta^2\Delta F^2 &= -\frac V2 \int\frac{d^dk}{(2\pi)^d}\,J.
\end{align}
The integrals over $k$ diverge unless we introduce a cutoff, which is tacitly
implied in Eq.~\eqref{deltaf}.

According to what we found so far for $n^-$, the free energy fluctuates with a
Gaussian distribution and a variance as given by Eq.~\eqref{deltaf}. This is,
we believe, the physically sensible solution: if, by the usual argument, a
spin glass sample is divided into many subsystems, each should contribute a
random dominant bulk term and a subdominant surface term to the free energy,
and by the central limit theorem this should give rise to a self-averaging,
Gaussian-distributed quantity.

If we repeat the above procedure for $n^+$, however, we find the following 
situation, which is illustrated most easily in a large-$k^2$ expansion of the 
propagators \cite{DeDominicisEtAl98},
\begin{multline}
    G_{\alpha\beta,\alpha\beta} = \frac{1}{k^2} + 
    \frac{2\tau + 2yq_{\alpha\beta}^2}{k^4} \\
    + \frac{1}{k^6}\left((2\tau+2yq_{\alpha\beta}^2)^2+
    w^2((q^2)_{\alpha\alpha}+(q^2)_{\beta\beta}-2q_{\alpha\beta}^2)\right) \\
    + \mathcal{O}(1/k^8).
\end{multline}
This expansion allows for a simple evaluation of $\sum_{\alpha<\beta}
G_{\alpha\beta,\alpha\beta}$, or rather $-n/2\int_{n}^1dx\,G^{xx}_{n}$, term
by term. While the first two terms are identical for $n^+$ and $n^-$, the
coefficient of the $1/k^6$-term is
\begin{multline}
    -\frac{n^+}{2}(4\tau^2-8y\tau \overline{q} +
    \frac{w^4x_1^4}{4y^2}(1-4x_{1}/5)) \\ 
    + \frac{{n^+}^2}{2}(4\tau^2-2w^2\overline{q})
    - {n^+}^6\frac{81w^4}{5\cdot128 y^2},
\end{multline}
while for $n^-$ one obtains the same but without a ${n^-}^6$-term. Here we
have used the abbreviation $\overline{q} = -\int_{n}^1 q^2_{n}(x)\,dx$, which
is independent of $n$. This shows that \textit{at least} one expansion
coefficient, and thus the physical consequences, changes when changing the
sign of $n$, since a nonzero $n^6$-term implies a non-gaussian probability
distribution of the free energy.

The difference between the two cases $n^+$ and $n^-$ can be eliminated by
modifying Parisi's original symmetry breaking scheme for $n^+$ in the
following way.  The $q$-matrix is divided into $p$ boxes on the diagonal, each
of which contains a Parisi-type symmetry broken matrix $Q\ab$, while the rest
of the matrix is zero. This is illustrated in Eq.~\eqref{qmatrix} for $p=4$,
\begin{align}
    \label{qmatrix}
q &=
\left(
\begin{array}{cccc}\cline{1-1}
\multicolumn{1}{|c|}{Q_{\alpha\beta}}&&& \\ \cline{1-2}
&\multicolumn{1}{|c|}{Q_{\alpha\beta}}&& \\ \cline{2-3}
&&\multicolumn{1}{|c|}{Q_{\alpha\beta}}& \\ \cline{3-4}
&&&\multicolumn{1}{|c|}{Q_{\alpha\beta}} \\ \cline{4-4}
\end{array}
\right).
\end{align}
Now we first let the number of symmetry breaking steps $R$ go to infinity,
then we let $p$ go to infinity, all the while keeping $n^+$ finite.

This procedure is in fact very closely related to, but certainly not identical
to, the original Parisi scheme.  If we introduce the usual ``integers''
$n^+=m_{0},m_{1},\dots,m_{R},m_{R+1}=1$ which characterise the block sizes of
the symmetry broken matrix, then the difference between this scheme and the
original one is that $m_{1}=n^+/p$ is allowed to go to 0 \textit{before}
$m_{0}$. This way of looking at it shows precisely where the difference
between positive and negative $n$ lies in the original scheme: for $n^-$,
$m_{1}$ \textit{does} go to 0 before $m_{0}$ since 0 now lies in the interval
$[n^-,1]$. It also shows that we may still use those exact results of
\cite{TemesvariEtAl94,DeDominicisEtAl98} for the modified scheme which were
derived for arbitrary $m$'s.

From Eq.~\eqref{rep3} it follows now that on mean-field level,
$\Hrep[q_{n^+}(x)]$ is to be replaced by $p\Hrep[q_{n^+/p}(x)]$ (there are $p$
Parisi blocks of size $n^+/p$ each), and in the limit $p\to\infty$ this kills
the ${n^+}^6$ and higher order terms in Kondor's solution (terms of order
${n^+}^m$ are replaced by $p({n^+}/p)^m\to 0\quad(p\to\infty)$ for $m>1$),
while only the term linear in $n^+$ survives, unaltered. Thus this procedure
has cured the inconsistency on the mean-field level without affecting the
average free energy itself. It also cures the discrepancy on the Gaussian
level, as the following argument shows.  The eigenvalue equation for the
Hessian which follows from Eq.~\eqref{rep3} is
\begin{equation}
(\lambda + 2\tau)f_{\alpha\beta} + 
w\sum_{\gamma}q_{\beta\gamma}f_{\gamma\alpha} +
w\sum_{\delta}q_{\delta\alpha}f_{\beta\delta} + 
2yq_{\alpha\beta}^2 f_{\alpha\beta}  = 0,
\end{equation}
where $\lambda$ is the eigenvalue and $f\ab$ is the eigenvector. Since $q\ab$
is zero in the off-diagonal blocks, the eigenvectors can be chosen to be zero
everywhere except in one block (and its counterpart on the opposite side of
the diagonal, if the block does not happen to be on the diagonal). If $f\ab$
is nonzero in a block on the diagonal, the problem reduces to the eigenvalue
problem of the original Parisi matrix, but of size $n^+/p$. There are $p$
blocks on the diagonal, i.e.\ we have to replace $-\frac{n^+}2
\int_{n^+}^1dx\, G^{xx}_{n^+}$ in Eq.~\eqref{propsum} by $-p\frac{n^+}{2p}
\int_{n^+/p}^1dx\, G^{xx}_{n^+/p}$, which again in the limit $p\to\infty$ kills
all terms of order ${n^+}^2$ and higher but leaves the linear term
unchanged. There are however additional eigenvectors which are nonzero in one
of the off-diagonal blocks, of which there are $p(p-1)/2$. According to
\cite{TemesvariEtAl94,DeDominicisEtAl98} the corresponding eigenvalues are 
again just the $x=0$ eigenvalues from the replicon sector, given by $1/f(s,t)$
from Eq.~\eqref{fdef}, and their contribution to $\partial I/\partial(k^2)$
can once more be reduced to the propagator $G^{00}_{n^+}$ and is $(p(p-1)/2)
(n^+/p)^2 G^{00}_{n^+/p} \to ({n^+}^2/2) G^{00}_{0}\quad(p\to\infty)$, while
higher order terms vanish. Collecting the pieces we get
\begin{align}
    \frac{\partial I}{\partial (k^2)} &=
    -\frac{n^+}{2} \int_{0}^1 dx\,G^{xx}_{0} + \frac{{n^+}^2}{2} G^{00}_{0}.
\end{align}
Comparison with Eq.~\eqref{propint} shows that this modified type of symmetry
breaking gives identical results to the ones derived before for $n^-$ and thus
eliminates the discrepancies between $n^+$ and $n^-$ limits.

The argument above also shows that the stability of the solution is not 
affected by the modification since all eigenvalues of the Hessian are the same 
as for the original scheme and are thus $\ge 0$.

We conclude with a few remarks. 
\begin{figure}[ht]
\epsfig{file=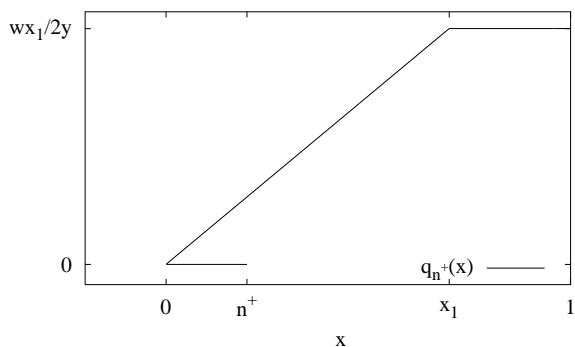,width=\columnwidth}
\caption{The multi-valued function $q_{n^+}(x)$ for the modified symmetry
breaking scheme. Note the similarity with $q_{n^-}(x)$ from
Fig.~\ref{qfunction}.}
\label{qmulti}
\end{figure}
A consequence of our modification of the symmetry breaking scheme is that the
Parisi $q$-function is not well defined for $n>0$ since the numbers
$m_0,m_1,\dots$ do not form a monotonous sequence any more. Instead, it turns
into the multi-valued function sketched in Fig.~\ref{qmulti}, elegantly
restoring symmetry with $q_{n^-}(x)$, which was violated in
Fig.~\ref{qfunction}. Integrals over functions $g$ of $q_{n^+}(x)$ have then
to be interpreted as
\begin{align}
\int_{n^+}^1 dx\,g(q_{n^+}(x)) &= 
\int_{n^+}^0 dx\,g(q_{n^+}^{\downarrow}(x)) + 
\int_0^1 dx\,g(q_{n^+}^{\uparrow}(x)),
\end{align}
where the arrows denote the branch in an obvious way.  Physical quantities
like the probability distribution of the order parameter, $P(q)=dx/dq$, are
unaffected by this man\oe uvre since they are only meaningful at $n=0$.

The need to modify Parisi's original symmetry breaking to the scheme described
here probably escaped notice as the two schemes become identical in the
limiting case $n=0$, which is usually considered.  Indeed higher order terms
in the expansion in $n$ have rarely been investigated for spin glasses as they
are extremely difficult to compute when using the established $n^+$-formalism
as in \cite{DeDominicisEtAl98}. With our modified scheme, however, (or,
equivalently, by working at $n<0$) not only are the higher order terms
guaranteed to be correct but as shown here they also become relatively
accessible. This may be important for future applications, e.g.\ for computing
interface energies in spin glasses (see \cite{AspelmeierEtAl02} for an attempt
in this direction). We would also expect that the modification of the Parisi
scheme described here is needed for the Sherrington-Kirkpatrick infinite range
model. Whether this is the case or not could be investigated from the sample
to sample fluctuations of the ground state energy of this model, which can be
determined numerically when the number of spins $N$ is small
\cite{PalassiniYoung02}.

\begin{acknowledgments}
T. A.\ acknowledges support by the German Academic Exchange Service (DAAD). We
thank C. De Dominicis and A. P. Young for many useful discussions and
M. M{\'e}zard for pointing out to us Ref.\ \cite{Kondor83}.
\end{acknowledgments}

\bibliography{Spinglass} 

\end{document}